# APLICAȚIE NUMERICĂ ȘI PROGRAM TURBO C AL METODEI GAUSS-JORDAN

**Cornelia Victoria ANGHEL DRUGĂRIN***

# NUMERICAL APPLICATION AND TURBO C PROGRAM USING THE GAUSS-JORDAN METHOD

**Abstract**. The article presents the general notions and algorithm about the Gauss-Jordan method. An eloquent example is given and the Turbo C program illustrated this method. We conclude that we can obtain by this method the determinant, by simple calculations and reducing the rounding errors

*Cuvinte cheie*: metoda Gauss-Jordan, matricea inversă, limbaj de programare Turbo C, pivot.

## 1. Noțiuni generale

Metoda eliminării în versiunea Gauss-Jordan, este o metodă de calcul a inversei unei matrice, care se rezumă la reducerea acesteia la matricea unitate, fapt datorită căruia se mai numește și metoda diagonalizării.

Această metodă se bazează pe o teoremă din algebra matriceală: *Dacă o matrice nesingulară $\underline{A}$ poate fi redusă la matricea unitate $\underline{I}$ prin înmulțirea la stânga cu un șir de matrice, atunci inversa $\underline{A}^{-1}$ se poate calcula prin înmulțirea lui $\underline{I}$ la stânga cu același șir de matrice în ordine inversă.*

## 2. Algoritmul metodei Gauss-Jordan

**Algoritmul de inversare** – cuprinde două procese de calcul derulate în paralel și are *n* pași:

I. Se fac inițializările :

$$\underline{A}^0 = \underline{A} \; ; \tag{1}$$

$$\underline{D}^0 = \underline{I} \; . \tag{2}$$

II. La pasul k, k = 1, 2, …, n, se calculează elementele matricelor $\underline{A}^k \, si \, \underline{D}^k$, utilizând formulele următoare:

$$a_{kj}^k = \frac{a_{kj}^{k-1}}{a_{kk}^{k-1}}, j = \overline{k+1, n} \tag{3}$$

$$d_{kj}^k = \frac{d_{kj}^{k-1}}{a_{kk}^{k-1}}, j = \overline{1, k} \tag{4}$$

$$a_{ii}^k = 1, \ i = \overline{1, k} \tag{5}$$

$$d_{ii}^k = 1, i = \overline{k+1, n} \tag{6}$$

$$a_{ij}^k = a_{ij}^{k-1} - a_{ik}^{k-1} \cdot a_{kj}^k, i = \overline{1, n}, i \neq k, j = \overline{k+1, n} \tag{7}$$

$$d_{ij}^k = d_{ij}^{k-1} - a_{ik}^{k-1} \cdot d_{kj}^k,, i = \overline{1, n}; \ i \neq k, j = \overline{1, k} \tag{8}$$

$$a_{ij}^k = 0, i = \overline{1, n}, i \neq j, j = \overline{1, k} \tag{9}$$

$$d_{ij}^k = 0, i = \overline{1, n}, i \neq j, j = \overline{k+1, n} \tag{10}$$

III În final se obţin matricea unitate, inversa şi valoarea determinantului:

$$\underline{I} = \underline{A}^n \tag{11}$$

$$\underline{A}^{-1} = \underline{D}^n \tag{12}$$

$$\det \underline{A} = a_{11}^0 \cdot a_{22}^1 \cdot a_{33}^2 \cdots a_{nn}^{n-1} \tag{13}$$

*Observaţii:*

1. Indicele superior corespunde pasului de calcul.

2. Din relaţiile (3) şi (4), deducem că, dacă elementul $a_{kk}^{k-1}$, numit **pivot** are valoare nulă sau modulul său este sub un *prag de zero* prestabilit, aşadar apar probleme la efectuarea împărţirilor. Pivotul nul nu implică neapărat că matricea este singulară. Trebuie încercate toate posibilităţile: poate fi adus in poziţia de pivot orice element $a_{ij}^{k-1}, i, j = \overline{k, n}$, cu schimbarea între ele a liniilor *k* şi *i* şi a coloanelor *k* şi *j*. Dacă toate aceste posibilităţi conduc la eşec, atunci vom spune că, matricea $\underline{A}$ este singulară.

3. În scopul reducerii erorilor de rotunjire se recomandă ca la fiecare pas să se aducă pe poziţia pivotului elementul de valoare absolut maximă, ales conform procedeului de la observaţia 2, procedeu numit **pivotare.** În final, trebuie efectuate din nou schimbarea de linii şi coloane.

4. Relaţiile (3) până la (10) sunt echivalente cu următoarele relaţii:

$$a_{kj}^{\,k} = \frac{a_{kj}^{\,k-1}}{a_{kk}^{\,k-1}}\;,\; d_{kj}^{\,k} = \frac{d_{kj}^{\,k-1}}{a_{kk}^{\,k-1}}\;,\;j=\overline{1,n}\;, \tag{14}$$

$$a_{ij}^{\,k} = a_{ij}^{\,k-1} + (-a_{ik}^{\,k-1})\cdot a_{kj}^{\,k}$$
$$d_{ij}^{\,k} = d_{ij}^{\,k-1} + (-a_{ik}^{\,k-1})\cdot a_{kj}^{\,k}\,,\; j=\overline{1,n}, i=\overline{1,n}, i\neq k, \tag{15}$$

cu interpretările :

- liniile corespunzătoare pivotului (liniile *k*) ale matricelor $\underline{A}^k$ si $\underline{D}^k$ se obțin prin împărțirea liniilor matricelor $\underline{A}^{k-1}, respectiv \underline{D}^{k-1}$, corespunzătoare pivotului, la pivot $(a_{kk}^{\,k-1})$ ;
- liniile i, cu $i=\overline{1,n}, i\neq k$, ale matricelor $\underline{A}^k si \underline{D}^k$ se obțin adunând la liniile i ale matricelor $\underline{A}^{k-1}, respectiv \underline{D}^{k-1}$, liniile *k* ale matricelor $\underline{A}^k si \underline{D}^k$ înmulțite cu $-\left(a_{ik}^{\,k-1}\right)$.

## 3. Program scris în aplicația TCLITE

Programul scris în limbajul Turbo C determină soluția pivotării /depivotării

```c
#include <stdio.h>
#include <conio.h>
#include <math.h>
int main()
{
 jordan();
 getche();
int n,k=0;
float x[10][10]; //matricea de lucru
float p[10]; //matrice de pivotare
float c[4][4]={{0,0,0,0},
          {0,1,1,1},  //matrice implicita
           {0,1,2,3},
           {0,1,3,6};
}
```

```c
void afis()
{ printf ("Iteratia k=%d \n",k);
    for(int j=1;j<=n;j++)
   { for(int i=1;i<=n;i++)
     printf("%2.3f ",x[i][j]);
     printf("\n");
    }
  printf("\n");
}
void copiez()
{//copiez matricea implicita in x si memorez ce mai mare val din col
float t;
 for(int j=1;j<=n;j++)
  {p[j]=0;
    for(int i=1;i<=n;i++)
     { t=c[i][j];
       x[i][j]=t;
       if (t>p[j]) p[j]=t;
     }
  }
} //sf copiez

void jordan()
{ float q,t;
  int i,j;
  printf("Pentru matricea implicita introduceti 0  \n");
  printf("n=");scanf("%d",&n);
  if(n!=0) {
         for(int j=1;j<=n;j++)
          { p[j]=0;
           for(int i=1;i<=n;i++)
             {printf("a[%d][%d]=",i,j);
              scanf("%f",&t);
              x[i][j]=t;
              if (t>p[j])p[j]=t; //calculez maxim
             } //sf for i
          } //sf for j
       } //sf if
  else {n=3;copiez();
       }
```

```c
//aici incepe algoritmul Gauss-Jordan
 printf("Matricea initiala \n");
 afis();
 for(int i=1;i<=n;i++)
  printf ("Maxim p[%d]=%2.5f ",i,p[i]);
 printf(" \n");
for(k=1;k<=n;k++)
{q=0;j=k;
    for(i=k;i<=n;i++)
      if(x[i][k]>q){q=x[i][k];j=i;}
 p[k]=j; //notez pivotul

 if(j!=k)
  {printf ("pivotez j=%d \n",j);
   for(int l=1;l<=n;l++)
         { q=x[j][l];
           x[j][l]=x[k][l];
           x[k][l]=q;}
  } //sf if

 q=x[k][k];
 x[k][k]=1;
 for(int j=1;j<=n;j++)
   x[k][j]=x[k][j]/q;
 for(int i=1;i<=n;i++)
  {if (i!=k)
    {q=x[i][k];
     x[i][k]=0;
     for(int j=1;j<=n;j++)
       x[i][j]=x[i][j]-x[k][j]*q;
    } //sf for sf if
  }
 afis();
} //sf bucla principala

//urmeaza operatia inversa , depivotarea

for(k=n-1;k>=1;k--)
 {
  j=p[k];
```

```
   if (j!=k)
    {
     printf("Depivotez linia %d \n",j);
     for(int i=1;i<=n;i++)
      { q=x[i][k];
        x[i][k]=x[i][j];
        x[i][j]=q;
      } //sf for i
    } //sf if
 } //sf for k
k=3;
afis();
} //sf jordan
```

## 4. Concluzii.

Principalul avantaj al metodei de calcul a inversei unei matrice în versiunea Gauss-Jordan este de a obține valoarea determinantului fără calcule laborioase, suplimentare. Scopul acesteia este reducerea erorilor de rotunjire recomandându-se ca la fiecare pas să se aducă pe poziția pivotului elementul de valoare absolut maximă, procedeu numit **pivotare.** La final, trebuie efectuate din nou schimbarea de linii și coloane în ordine inversă.

## 5. Bibliografie

* Ș.l..dr.ing. Cornelia Victoria ANGHEL DRUGĂRIN, Facultatea de Inginerie Electrică și Informatică, Universitatea „Eftimie Murgu" din Reșița, membru AGIR, cenzor filiala Caraș-Severin, c.anghel@uem.ro .